\documentclass{JHEP}
\usepackage{amsmath,amssymb,mathrsfs,dsfont,cite}
\usepackage[all]{xy}

\numberwithin{equation}{subsection}
\newcommand{\be}{\begin{equation}}
\newcommand{\ee}{\end{equation}}
\newcommand{\bea}{\begin{eqnarray}}
\newcommand{\eea}{\end{eqnarray}}
\newcommand{\kab}{|k_1,k_2\rangle}
\newcommand{\kba}{|k_2,k_1\rangle}

\title{\begin{flushright}
 \small SU-4252-860
\end{flushright}Twisted Poincar\'{e} Invariant Quantum Field
Theories}

\author{A. P. Balachandran\\ Department of Physics, Syracuse University,
Syracuse NY, 13244-1130, USA. \\ E-mail: \email{bal@phy.syr.edu}}
\author{A. Pinzul\\ Instituto de F\'{i}sica, Universidade de S\~{a}o Paulo
C.P. 66318, S˜ao Paulo, SP, 05315-970, Brazil \\ E-mail:
\email{apinzul@fma.if.usp.br}}
\author{B. A. Qureshi\\ Department of Physics, Syracuse University,
Syracuse NY, 13244-1130, USA. \\ E-mail:
\email{bqureshi@phy.syr.edu}}

\abstract{ It is by now well known that the Poincar\'e group acts
on the Moyal plane with a twisted coproduct. Poincar\'e invariant
classical field theories can be formulated for this twisted
coproduct. In this paper we systematically study such a twisted
Poincar\'e action in quantum theories on the Moyal plane. We
develop quantum field theories invariant under the twisted action
from the representations of the Poincar\'e group, ensuring also
the invariance of the $S$-matrix under the twisted action of the
group . A significant new contribution here is the construction of
the Poincar\'e generators using quantum fields.}

\begin{document}




\section{Introduction}
Recently it was pointed out that the apparent lack of Poincar\'e
symmetry on the Groenewald-Moyal (GM) plane can be given a new
interpretation, using the known results in quantum group theory,
according to which the Poincar\'e symmetry is still preserved,
though with a new coproduct\cite{aschieri1,aschieri2,chaichian}.
For example, consider the following integral on the GM plane:
\begin{align}
S[\phi]\,&=\,\int
d^dx\,\phi(x)\ast\phi(x)\ast\cdots\ast\phi(x),\label{func}\\
\phi\ast\phi(x)\,&=\,
\phi\,e^{\frac{i}{2}\theta^{\mu\nu}{\overleftarrow
\partial_\mu}{\overrightarrow \partial_\nu}}\,\phi(x),\label{star1}
\end{align}
where $\phi$ is a scalar field and the $\ast-$product as defined in
(\ref{star1}) is used to multiply functions on the GM plane. The
fields carry a representation of the Poincar\'e group and transform
under a Poincar\'e tranformation $g$ in the usual way:
\begin{equation}
g\triangleright\phi(x)\ =\ \phi(g^{-1}x)\ .
\end{equation}

If we treat the  product $\phi\ast\phi\ast\cdots\ast\phi$  again
as a scalar function and  transform it in the usual way, the
integral $S$ will be invariant. But the problem is that if we
transform the fields individually in (\ref{func}), then
\begin{equation}
g\triangleright\phi\,\ast\,g\triangleright\phi\,\ast\cdots\ast\,g\triangleright\phi\,\neq
\,g\triangleright\,(\ \phi\ast\phi\ast\cdots\ast\phi\ ).\label{it}
\end{equation}
 Hence $S$  is not Poincar\'e invariant in the sense that
\begin{equation}
S[g\triangleright\phi]\neq \ S[\phi]\ .
\end{equation}

But the group action on individual fields in a product comes from a
coproduct $\Delta_0$ on the group. The non-invariance of $S$ occurs
because the usual coproduct $\Delta_0$ on the Poincar\'e group,
$$\Delta_0(g)=g\otimes
g$$
 is not compatible with $\ast-$multiplication.

 Now there is a well
defined way of deforming the coproduct so that the deformed
coproduct $\Delta_\theta$  gives us the right hand side in
(\ref{it}), that is,
\begin{equation}
m_\theta\,\Delta_\theta(g)\, f\otimes g\ =\
g\triangleright\,m_\theta\,(f\otimes g) \label{comp}
\end{equation}
where $m_\theta$ is the $\ast$-multiplication map:

\begin{equation}
m_\theta(f\otimes g)=f\ast g\ .
\end{equation}
We recall this coproduct in Section 2.

The invariance of classical field theories for the Poincar\'e group
action with the twisted coproduct does not automatically mean that
 quantum theories associated with such classical field theories will also be
invariant under the twisted Poincar\'e transformations. For example,
the Poincar\'e invariance of the measure used to define the
functional integral has to be reconsidered \cite{stat,oeckl}.

In this paper we take the route of Hamiltonian quantum theory to
study quantum theories. We examine the construction of  quantum
field theories with  Poincar\'e invariance with the twisted
coproduct (for the Poincar\'e group) as the symmetry principle.
Our treatment closely follows the general theory of quantum group
symmetries in quantum mechanics as discussed by Mack and
Schomerus\cite{mack1,mack2}.

The paper is organized as follows. We briefly discuss the
Drinfel'd twist of the coproduct of the Poincar\'e group in
Section \ref{stgp}. Section \ref{qft} reviews the statistical
properties of the Hilbert space of a quantum  theory with the Hopf
algebra associated with the twisted coproduct as its symmetry and
its well-known connection with the $\mathcal{R}$ matrix of the
Hopf algebra is explained. In section \ref{qg} explicit
expressions for the quantum generators for the Lie algebra of the
Poincar\'e group (with the twisted coproduct) are given in terms
of creation and annihilation operators . Section \ref{ih}
discusses the form of interaction Hamiltonians which give us
Lorentz invariant $S$-matrices. Section \ref{sdef} discusses some
conventions in the definition of creation/annihilation operators.
Section \ref{scon} concludes the paper.

\section{The Twisted Coproduct of the Poincar\'e Group}\label{stgp}
For completeness and fixing the notation, we briefly recall the
Drinfel'd twist of the coproduct on the Poincar\'e group. For
details see\cite{chaichian}.

The usual Poincar\'e group $P$ has associated with it a canonical
coproduct $\Delta_0$,
\begin{equation}
\Delta_0(g)\ =\ g\,\otimes\,g\ ,\ \ g\in P\label{coprod}
\end{equation}
or at the Lie algebra level, for $u$ in the Lie algebra
$\mathcal{P}$,
\begin{equation}
\Delta_0(u)\ =\ u\,\otimes\,\mathds{1}\,+\,\mathds{1}\,\otimes\,u.
\end{equation}
These definitions extend to the group algebra $\mathfrak{P}$ of
$P$ and the universal enveloping algebra $\mathscr{P}$ of
$\mathcal{P}$ by linearity. The group algebra $\mathfrak{P}$ and
the universal enveloping algebra $\mathscr{P}$ have the full Hopf
algebra structure with the following counit  and antipode defined
on the pure group elements $g$ of $\mathfrak{P}$ by
\begin{equation}
\epsilon(g)\,=\,\mathds{1}\ \ ,\ \ S(g)\,=\,g^{-1}\label{def2}
\end{equation}
which is then extended by linearity to the whole $\mathfrak{P}$.
Alternatively on the generators $u$ and $\mathds{1}$ of
$\mathscr{P}$, $\epsilon$ and $S$ read
\begin{eqnarray}
\epsilon(u)\,=\,0\ &,&\ \epsilon(\mathds{1})\,=\,\mathds{1}\nonumber\\
S(u)\,=\,-u\ &,&\ S(\mathds{1})\,=\,\mathds{1}\label{def1}\ .
\end{eqnarray}
They are then defined on all of $\mathscr{P}$ by linearity.

We can think of the elements of the universal enveloping algebra
$\mathscr{P}$ as living in the group algebra $\mathfrak{P}$ and
vice versa. Notice that we can either define the above structures
on the pure group elements and then by linearity and limiting
procedures, this defines them on the Lie algebra, or we can have
the definitions of $\Delta$, $S$ and $\epsilon$  on the Lie
algebra elements and by linearity they induce the coproduct etc.
on pure group elements. So only one of the definitions in
(\ref{def1}, \ref{def2}) is really needed.

We can define a new coproduct $\Delta_\theta$ on $\mathfrak{P}$
and $\mathscr{P}$ by the Drinfel'd twist:
\begin{eqnarray}
\Delta_\theta(g)\ &=&\
F_{\theta}^{-1}\,\Delta_0(g)\,F_{\theta}\label{tcop1}\\
\Delta_\theta(u)\ &=&\
F_{\theta}^{-1}\,\Delta_0(u)\,F_{\theta},\label{tcop}
\end{eqnarray}
where
\begin{equation}
F_{\theta}\,=\,e^{-\frac{i}{2}\theta_{\mu\nu}P^\mu\otimes P^\nu}\
\ \ , \ \ \
F_{\theta}^{-1}\,=\,e^{\frac{i}{2}\theta_{\mu\nu}P^\mu\otimes
P^\nu} \label{twist}
\end{equation}
where $P^\mu$ is momentum operator.

With this coproduct (and the same counit and antipode as before), we
obtain the twisted Hopf algebra of the Poincar\'e group.

The Poincar\'e group with the twisted coproduct acts on the algebra
$\mathcal{A}_\theta$ of functions $f\,\in\,\mathbb{R}^d$ with the
product defined through a $\ast-$product, compatibly in the sense of
eq. (\ref{comp}). The $\ast-$product is given by
\begin{equation}
m_\theta (\phi\otimes \psi )= m_0 \mathcal{F_{\theta}}\,
\phi\otimes\psi. \label{star}
\end{equation}
where $\mathcal{F_{\theta}}$ is the differential operator
representing $F_{\theta}$ on the space of functions and $m_0$ is the
usual untwisted multiplication map.

\section{Twisted Coproduct and the Hilbert Space}\label{qft}
Here we review the construction of the Hilbert space with the
twisted Hopf-Poincar\'e symmetry, using the language of Mack and
Schomerus. The general idea of twisted symmetries in the quantum
Hilbert space is given in \cite{mack1, mack2}, while most of the
results of this section have been given in \cite{stat}.

\subsection{Transformation Law for Hilbert Space}
The Hilbert space of quantum fields consists of all multi-particle
states. We start with the single particle states.
\subsubsection{Single Particle States}
As usual we identify the single particle states with the
one-particle irreducible representations of the (identity
component of) Poincar\'e group. For simplicity, we consider the
massive spinless case so that a basis of the states can be labeled
just by momenta.

We choose the following normalization for the single particle
states:
\begin{align}
\langle k|p\rangle\ &=\ 2 k^0\delta^3(k-p)\ ,\\
k^0 &=\sqrt{{\overrightarrow k}^2+m^2}\ ,\nonumber\\
m &=\textrm{mass of the particle.}\nonumber
\end{align}

We have unitary  operators $U(g)$ on the Hilbert space which form
a representation of the Poincar\'e group ${P}_+^{\uparrow}$. On
single particle states, they act as usual,
\begin{equation}
U(g)|k\rangle\ =\ |gk\rangle\ \ \ ,\ g\in P\label{gaction}
\end{equation}
and hence
\begin{equation}
U(g_1)U(g_2)\ =\ U(g_1g_2)\ \ \ \ \ \textrm{on single particle
states.}
\end{equation}

We can also write eq (\ref{gaction}) as
\begin{equation}
U(g)|k\rangle\ =\ \int
\frac{d^3k'}{2k'^{0}}\,\rho_{k'k}(g)|k'\rangle
\end{equation}
where
\begin{eqnarray}
\rho_{k'k}(g)&=&\langle k'|U(g)|k\rangle\nonumber\\
&=&2k'^0\delta^3(k'-gk).
\end{eqnarray}
(The integral(sum) will always be assumed over the repeated indices,
unless otherwise stated, though we will not explicitly write it.)
$\rho_{k'k}(g)$ are the matrix elements of $U(g)$ and form a
representation of the group:
\begin{equation}
\rho_{kp}(g_1)\rho_{pk'}(g_2)\ =\ \rho_{kk'}(g_1g_2).
\end{equation}

The representation of the group extends naturally to the group
algebra by linearity. For $f=\sum_i a_i g_i\,\in\ \mathfrak{P}$
and $g_i\,\in\,P$
\begin{equation}
U(f)=\sum_i a_i U(g_i)\ .
\end{equation}
(More generally we should write $f=\int dg\,f(g)\,g$ and
$U(f)=\int dg\,f(g)\,U(g)$, where $dg$ is the Haar measure on the
Poincar\'e group, but it does not change any of the arguments.)

 Again we can write
\begin{equation}
U(f)|k\rangle\ =\ \rho_{k'k}(f)|k'\rangle
\end{equation}
where
\begin{equation}
\rho(f)\ =\ \sum_i a_i \rho(g_i)\ .
\end{equation}
It is straightforward to check that $U(f)$ and the matrix
$\rho(f)$ form a representation of $\mathfrak{P}$.

Here it is worth noting that we define the vacuum to be invariant
under the group so that under a pure group element,
\begin{equation}
U(g) |0\rangle\ =\ |0\rangle
\end{equation}
but under a general group algebra element $f$,
\begin{align}
U(f)|0\rangle &=\sum_i a_iU(g_i)|0\rangle \nonumber\\
&=\sum_i a_i|0\rangle \nonumber\\
&=\epsilon(f)|0\rangle\ .
\end{align}

The single particle sector is unaffected by the twist. The twist
has only changed the coproduct and  the coproduct does not show up
in the single particle representations.

\subsubsection{Multi-Particle States}

Let us for simplicity first consider the two-particle sector. The
two-particle states carry two momentum indices and hence transform
according to the two-fold tensor product of single particle
representations. The tensor products of representations are
defined using the coproduct. In the undeformed case, the coproduct
over the group is given by eq (\ref{coprod}), so we have the
familiar transformation, for a pure group element $g$,
\begin{eqnarray}
U(g)|k_1,k_2\rangle \ &=&\
\rho_{k_1'k_1}\,\otimes\,\rho_{k_2'k_2}\,(\Delta_0(g))\,|k_1',k_2'\rangle \\
&=&\ \rho_{k_1'k_1}\,\otimes\,\rho_{k_2'k_2}\,(g\otimes g)\,|k_1',k_2'\rangle \\
&=&\ \rho_{k_1'k_1}(g)\,\rho_{k_2'k_2}(g)\,|k_1',k_2'\rangle \\
&=&\ |gk_1,gk_2\rangle \ .
\end{eqnarray}

But in the twisted case, we must use the deformed coproduct
$\Delta_\theta$, so that

\begin{eqnarray}
U(g)|k_1,k_2\rangle \ &=&\
\rho_{k_1'k_1}\,\otimes\,\rho_{k_2'k_2}\,(\Delta_\theta(g))\,|k_1',k_2'\rangle \nonumber\\
&=&\ \rho_{k_1'k_1}\,\otimes\,\rho_{k_2'k_2}\
(F_{\theta}^{-1}\,(g\otimes g)\,
F_{\theta})\ |k_1',k_2'\rangle \nonumber\\
&=&\ e^{-\frac{i}{2}\theta_{\mu\nu}k_1^\mu
k_2^\nu}\,e^{\frac{i}{2}\theta_{\mu\nu}(gk_1)^\mu (gk_2)^\nu}\,
|gk_1,gk_2\rangle .\label{twcop1}
\end{eqnarray}

Similarly for an $n$-particle state we have the transformation law
\begin{eqnarray}
\lefteqn{U(g)|k_1\,,\,k_2\,,\,\cdots\,k_n\rangle \ =}\nonumber\\
&&\rho_{k_1'k_1}\,\otimes\,\rho_{k_2'k_2}\,\otimes\,\cdots\,\otimes\,\rho_{k_n'k_n}\,\nonumber\\
&&\{(\mathds{1}\,\otimes\,\cdots\,\otimes\,\mathds{1}\,\otimes\,\Delta_\theta)\,\cdots\,
(\mathds{1}\,\otimes\,\Delta_\theta)\,\Delta_\theta(g)\}\,|k_1'\,,\,k_2'\,,\,\cdots\,k_n'\rangle
\ .\label{twcop}
\end{eqnarray}
This equation carries the main content of twisting.

Clearly $U(g)$ also gives a representation of the twisted Hopf
Algebra $\mathfrak{P}$.

\subsection{Statistics of States}

Let $|k_1, k_2\rangle_{s_0,a_0} $ denote a two-particle boson
(fermion) state with definite momenta for each particle for
$\theta_{\mu\nu}=0$. For identical particles the state
$|k_2,k_1\rangle_{s_0,a_0}$ is not an independent state and is
related to $|k_1,k_2\rangle_{s_0,a_0}$, in the usual case, by the
equivalence
\begin{equation}
|k_1,k_2\rangle_{s_0,a_0} \sim |k_2,k_1\rangle_{s_0,a_0}
\label{ueq}
\end{equation}
since
\begin{eqnarray}
& &|k_2,k_1\rangle_{s_0,a_0} = \frac{1}{2}(|k_2,k_1\rangle \pm
|k_1,k_2\rangle ) = \pm |k_1,k_2\rangle_{s_0,a_0} \ ,  \nonumber\\
& &|k_1,k_2\rangle := |k_1\rangle \otimes |k_2\rangle\ . \nonumber
\end{eqnarray}

A statistics operator $\tau$ takes one vector of the equivalence
class to the other. Given the statistics operator $\tau$ the vectors
in the Hilbert space are identified by
\begin{equation}
V=\tau V \ .\label{statt}
\end{equation}

The usual statistics operator $\tau_0$ corresponding to
(\ref{ueq}) is just the flip operator
\begin{equation}
\tau_0|k_1,k_2\rangle = |k_2,k_1\rangle
\end{equation}
with
\begin{equation}
|k_1,k_2\rangle_{s_0,a_0} = \frac{1}{2}(\mathds{1} \pm \tau_0
)|k_1,k_2\rangle \ . \nonumber
\end{equation}

Compatibility of Lorentz invariance with statistics means that two
vectors, $V$ and $W$, in the Hilbert space which are identified by
an equivalence relation   should transform under the group action,
to vectors which are again identified with each other according to
the same relations. In other words, an equivalence class should
transform into an equivalence class. Applying this to two-particle
sector (with an obvious generalization to arbitrary number of
particles) one can easily see that this is equivalent to the
requirement that statistics operator defined by Eq.(\ref{statt})
commutes with the coproduct.

The immediate consequence of this is that the `commutative'
statistics operator $\tau_0$ leads to statistics not compatible
with the twisted Poincar\'{e} symmetry. This is due to the fact
that the twisted coproduct $\Delta_\theta$ is not cocommutative,
i.e.
\begin{equation}
\tau_0\Delta_\theta(g)=\tau_0(g_\alpha^{(1)}\otimes
g_\alpha^{(2)})\ = (g_\alpha^{(2)}\otimes g_\alpha^{(1)})\tau_0\ =
\ \Delta_\theta'(g)\tau_0 \ \neq\ \Delta_\theta(g)\tau_0
\label{sweedler}
\end{equation}
where we have written $\Delta_\theta (g)$ in the Sweedler notation
with a summation over $\alpha$.

But using the operator of the twist, Eq.(\ref{twist}), and the
definition of the twisted coproduct, Eq.(\ref{twcop1}), one can
easily construct an appropriate deformation of the statistics
operator. An evident solution, $\tau_\theta$, that commutes with
$\Delta_\theta$ is
$$
\tau_\theta = F^{-1}_\theta \tau_0 F_\theta \ .
$$

Note that
$$
\tau_\theta =\tau_0 F_\theta^2 = \tau_0 \mathcal{R}\ ,
$$
where $\mathcal{R} := F_\theta^2$ satisfies the following identity
\begin{equation}
\mathcal{R}\ \Delta_\theta\ =\ \Delta'_\theta\
\mathcal{R}\label{R} \nonumber\ .
\end{equation}
$\mathcal{R}$ is called $R$-matrix of the twisted Poincar\'{e}
group.

The two-particle momentum eigenstates with twisted statistics are
thus \footnote{We use the same notation, $\tau_\theta$, for the
statistics operator and its representation on the Hilbert space.}
\begin{equation}
|k_1,k_2\rangle_{s_\theta,a_\theta}=\frac{\mathds{1}\pm
\tau_\theta}{2}|k_1,k_2\rangle \ ,
\end{equation}
where we have to use a representation of $F_\theta$ on the Hilbert
space. We denote it by $\mathcal{F}_\theta$ (cf. Eq.(\ref{star})):
$$
\mathcal{F}_\theta |k_1,k_2\rangle =
e^{-\frac{i}{2}\theta_{\mu\nu}k^\mu_1 k^\nu_2}|k_1,k_2\rangle \ .
$$
It follows that
\begin{equation}
\kab_{s_\theta,a_\theta}\ =\ \pm e^{i\theta_{\mu\nu}k_2^\mu
k_1^\nu}\,\kba_{s_\theta,a_\theta} \ .
\end{equation}

One can easily generalize this to a multi-particle state. Consider
a three-particle state. Again comparing the transformation of the
state $|k_1,k_2,k_3\rangle_{s_\theta,a_\theta} $ and
$|k_3,k_2,k_1\rangle_{s_\theta,a_\theta} $ one finds out that they
must be related by

\begin{equation}
|k_1,k_2,k_3\rangle_{s_\theta,a_\theta} \ =\pm
e^{i\theta_{\mu\nu}k_3^\mu k_2^\nu}e^{i\theta_{\mu\nu}k_3^\mu
k_1^\nu}e^{i\theta_{\mu\nu}k_2^\mu k_1^\nu}\
|k_3',k_2',k_1'\rangle_{s_\theta,a_\theta} \ . \label{tstat}
\end{equation}

So when we exchange two particles, we must commute the two labels
past the neighboring labels to bring them to the desired place,
and include the exponential factors as in (\ref{tstat}) for each
such permutation.

Generalizing $\tau_\theta$ to transpositions of adjacent particles
in $|k_1\rangle \otimes |k_2\rangle \otimes |k_3\rangle \equiv
|k_1,k_2,k_3\rangle$ and using them to fully symmetrize
(antisymmetrize) $|k_1,k_2,k_3\rangle$, we can explicitly write
$|k_1,k_2,k_3\rangle_{s_\theta,a_\theta}$.

\subsection{Scalar Product}\label{scsec}
Now we must chose a scalar product for multi-particle states which
is compatible with the above statistics (and hence in turn
compatible with the Lorentz group in the sense that the operators
$U(g)$ are unitary in this scalar product).

First consider the two-particle states. A scalar product compatible
with the above statistics (upto multiplicative constants) is
\begin{eqnarray}
\lefteqn{(|k_1,k_2\rangle_{s_\theta,a_\theta} \ ,\ |p_1,p_2\rangle_{s_\theta,a_\theta} )  =}\nonumber\\
&&4k_1^0 k_2^0\,[\ \delta^3(k_1-p_1)\delta^3(k_2-p_2)
 \pm\ e^{-i\theta_{\mu\nu}p_2^\mu
p_1^\nu}\delta^3(k_1-p_2)\delta^3(k_2-p_1)\ ]\ .\label{scalar}
\end{eqnarray}
Now the right-hand side has the same symmetry properties as the
left-hand side. Note that the phase factor which gives the correct
symmetry can only be included in the second term because of the
positivity of scalar product. Also this single phase factor gives
the correct symmetry under the exchange of particles in both
vectors.

Generalization to higher number of particles is simple. The first
term which involves delta functions in the same ordering as in the
vectors in the scalar products is without any phase factor. The
delta functions in the other terms can all be obtained from
permutations of the second momentum labels in the first term, and we
include a phase factor like in (\ref{scalar}), for each such
permutation. Again as in (\ref{tstat}), when we permute two labels
which are not nearest neighbors, we must commute them past the
neighbors and include the phase factor for each of such permutation.
As an illustration we write the scalar product of three particle
states.
\begin{flalign}
(|p_1,p_2,p_3\rangle_{s_\theta,a_\theta}
,&|k_1,k_2,k_3\rangle_{s_\theta,a_\theta} )\ =\ 2^3 p_1^0 p_2^0
p_3^0
[\delta^3(p_1-k_1)\delta^3(p_2-k_2)\delta^3(p_3-k_3)\nonumber\\
&\pm e^{-i\theta_{\mu\nu}k_3^\mu
k_2^\nu}\delta^3(p_1-k_1)\delta^3(p_2-k_3)\delta^3(p_3-k_2)\nonumber\\
&\pm e^{-i\theta_{\mu\nu}k_2^\mu
k_1^\nu}\delta^3(p_1-k_2)\delta^3(p_2-k_1)\delta^3(p_3-k_3)\nonumber\\
&+e^{-i\theta_{\mu\nu}k_2^\mu k_1^\nu}e^{-i\theta_{\mu\nu}k_3^\mu
k_1^\nu}\delta^3(p_1-k_2)\delta^3(p_2-k_3)\delta^3(p_3-k_1)\nonumber\\
&+e^{-i\theta_{\mu\nu}k_3^\mu k_1^\nu}e^{-i\theta_{\mu\nu}k_3^\mu
k_2^\nu}\delta^3(p_1-k_3)\delta^3(p_2-k_1)\delta^3(p_3-k_2)\nonumber\\
&\pm e^{-i\theta_{\mu\nu}k_3^\mu
k_1^\nu}e^{-i\theta_{\mu\nu}k_3^\mu
k_2^\nu}e^{-i\theta_{\mu\nu}k_2^\mu
k_1^\nu}\delta^3(p_1-k_3)\delta^3(p_2-k_2)\delta^3(p_3-k_1)]\ .
\end{flalign}

Now that we know the scalar product we can define a vector
$\langle k_1,k_2|$ dual to $|k_1,k_2\rangle $. Note that from the
definition of the scalar product, the phase in the statistics of
the dual vector has negative sign  relative to the vector. i.e.,
\begin{equation}
_{s_\theta,a_\theta}\langle k_1,k_2|\ =\pm
e^{-i\theta_{\mu\nu}k_2^\mu k_1^\nu}{ _{s_\theta,a_\theta}}\langle
k_2,k_1| \ .
\end{equation}

\section{Quantum Generators  for the Poincar\'e
Group with Twisted  Coproduct}\label{qg}
 We now give the explicit formulae for the quantum
operators for the twisted Poincar\'e group in terms of creation and
annihilation operators.

Hereafter in this paper, we focus on spin zero particles and
fields.

\subsection{Creation/Annihilation Operators}
We define the creation operator $a_k^\dagger$ to be the operator
which adds a particle with momentum $k$ to the list of particles as
usual. But there is an ambiguity as to whether the particle should
be added to the left or to the right of the list. In the usual case
the left and the right of the list has no inherent meaning because
the two states got by either of the procedures transform in a
similar manner under the Poincar\'e group and have the same symmetry
properties with respect to the exchange of the new particle with any
of the others already present. But now this is no longer the case.
The two states transform differently under the twisted action of the
Poincar\'e group (because of the non-cocommutativity of the
coproduct). Hence we must make a choice.

We chose to define $a_k^{\dagger}$ to be the operator which adds a
particle to the right of the particle list
\begin{equation}
a_{k}^{\dagger}|k_1,k_2,\cdots,k_n\rangle_{s_\theta} \ =\
|k_1,k_2,\cdots,k_n,k\rangle_{s_\theta}\ .
\end{equation}
(Later we will discuss what would change if we define $a_k^\dagger$
the other way).

Applying $a_k^\dagger$ twice and using the statistics we can
easily see that the $a_k^\dagger$'s have commutation relation
\begin{equation}
a_{k_2}^\dagger a_{k_1}^\dagger\ =\ e^{i\theta_{\mu\nu}k_2^\mu
k_1^\nu}\,a_{k_1}^\dagger a_{k_2}^\dagger \ .\label{adcom}
\end{equation}

Now let us see what is the effect of $a_k$, the adjoint of
$a_k^\dagger$, on a general state. As usual we can find it out by
calculating the matrix elements of
$a_k|k_1,k_2,\cdots,k_n\rangle_{s_\theta} $ with other states. We
have
\begin{align}
(|p_1,p_2,\cdots,p_{n-1}\rangle_{s_\theta} &,a_k|k_1,k_2,\cdots,k_{n-1},k_n\rangle_{s_\theta} )\nonumber\\
&=(a^\dagger_k|p_1,p_2,\cdots,p_{n-1}\rangle_{s_\theta} ,|k_1,k_2,\cdots,k_{n-1},k_n\rangle_{s_\theta} )\nonumber\\
&= (|p_1,p_2,\cdots,p_{n-1},k\rangle_{s_\theta}
,|k_1,k_2,\cdots,k_{n-1},k_n\rangle_{s_\theta} )\ .
\end{align}
But using the scalar product in section (\ref{scsec}), it is
easily seen that this matrix element is the same as that of
$|p_1,p_2,\cdots,p_{n-1}\rangle_{s_\theta} $ with the state
\begin{multline}
2k^0\delta^3(k-k_n)|k_1,k_2,\cdots,k_{n-1}\rangle_{s_\theta}
+2k^0e^{-i\theta_{\mu\nu}k^\mu
k_n^\nu}\delta^3(k-k_{n-1})|k_1,k_2,\cdots,k_{n-2},k_n\rangle_{s_\theta} \\
+\cdots\ 2k^0 e^{-i\theta_{\mu\nu}k^\mu
(k_n+k_{n-1}+\cdots+k_2)^\nu}\delta^3(k-k_1)|k_2,k_3,\cdots,k_n\rangle_{s_\theta}\
. \nonumber
\end{multline}
Since $|p_1,p_2,\cdots,p_{n-1}\rangle_{s_\theta} $ is a general
state, we have that
\begin{align}
a_k|k_1,k_2,\cdots,&k_{n-1},k_n\rangle_{s_\theta} =\nonumber\\
&2k^0\delta^3(k-k_n)|k_1,k_2,\cdots,k_{n-1}\rangle_{s_\theta} \nonumber\\
&+2k^0e^{-i\theta_{\mu\nu}k^\mu
k_n^\nu}\delta^3(k-k_{n-1})|k_1,k_2,\cdots,k_{n-2},k_n\rangle_{s_\theta} +\cdots\nonumber\\
&\cdots +2k^0 e^{-i\theta_{\mu\nu}k^\mu
(k_n+k_{n-1}+\cdots+k_2)^\nu}\delta^3(k-k_1)|k_2,k_3,\cdots,k_n\rangle_{s_\theta}\
.
\end{align}

The commutator of $a_k$ can be directly found by taking the adjoint
of (\ref{adcom}) , and we find
\begin{equation}
a_{k_2} a_{k_1}\ =\ e^{i\theta_{\mu\nu}k_2^\mu k_1^\nu}\,a_{k_1}
a_{k_2}\ .\label{com}
\end{equation}

Now let us find the commutator of $a_k$ and $a_k^\dagger$. Acting
on a general state $|p_1,p_2,\cdots,p_n\rangle_{s_\theta} $ by
$a^\dagger_{k_2}a_{k_1}$, we get
\begin{align}
a^\dagger_{k_2}a_{k_1}&|p_1,p_2,\cdots,p_n\rangle_{s_\theta} =\nonumber\\
&2k_1^0\delta^3(k_1-p_n)|p_1,p_2,\cdots,p_{n-1},k_2\rangle_{s_\theta} \nonumber\\
&+2k_1^0e^{-i\theta_{\mu\nu}k_1^\mu
p_n^\nu}\delta^3(k_1-p_{n-1})|p_1,p_2,\cdots,p_{n-2},p_n,k_2\rangle_{s_\theta} +\cdots\nonumber\\
&\cdots +2k_1^0 e^{-i\theta_{\mu\nu}k_1^\mu
(p_n+p_{n-1}+\cdots+p_2)^\nu}\delta^3(k_1-p_1)|p_2,p_3,\cdots,p_n,k_2\rangle_{s_\theta}
\label{ada}\ .
\end{align}
On the other hand acting by $a_{k_1}a^\dagger_{k_2}$ gives us
\begin{align}
a_{k_1}a^\dagger_{k_2}&|p_1,p_2,\cdots,p_n\rangle_{s_\theta} =\nonumber\\
&2k_1^0\delta^3(k_1-k_2)|p_1,p_2,\cdots,p_{n-1},p_n\rangle_{s_\theta} \nonumber\\
&+2k_1^0e^{-i\theta_{\mu\nu}k_1^\mu
k_2^\nu}\delta^3(k_1-p_n)|p_1,p_2,\cdots,p_{n-1},k_2\rangle_{s_\theta} +\cdots\nonumber\\
&+2k_1^0e^{-i\theta_{\mu\nu}k_1^\mu
(p_n+k_2)^\nu}\delta^3(k_1-p_{n-1})|p_1,p_2,\cdots,p_{n-2},p_n,k_2\rangle_{s_\theta} +\cdots\nonumber\\
&\cdots +2k_1^0 e^{-i\theta_{\mu\nu}k_1^\mu
(p_n+p_{n-1}+\cdots+p_2+k_2)^\nu}\delta^3(k_1-p_1)|p_2,p_3,\cdots,p_n,k_2\rangle_{s_\theta}
\label{aad}\ .
\end{align}
Dividing both sides of (\ref{aad}) by $e^{-i\theta_{\mu\nu}k_1^\mu
k_2^\nu}$ and subtracting from (\ref{ada}), we get
\begin{equation}
a^\dagger_{k_2}a_{k_1}\ =\ e^{i\theta_{\mu\nu}k_1^\mu
k_2^\nu}\,a_{k_1}a^\dagger_{k_2}\,-\,2k_1^0\delta^3(k_1-k_2)\
.\label{adacom}
\end{equation}
Or
\begin{equation}
a_{k_2}a^\dagger_{k_1}\ =\ e^{i\theta_{\mu\nu}k_1^\mu
k_2^\nu}\,a^\dagger_{k_1}a_{k_2}\ +\ 2k_1^0\delta^3(k_1-k_2)\ .
\end{equation}

\subsection{Transformation Law for Creation/Annihilation Operators}

The transformation law for creation and annihilation operators is
deduced from the transformation law of states. According to
(\ref{twcop1}), we want to have

\begin{equation}
U(g)a^\dagger_{k_2}a^\dagger_{k_1}|0\rangle =
e^{-\frac{i}{2}\theta_{\mu\nu}k_1^\mu
k_2^\nu}\,e^{\frac{i}{2}\theta_{\mu\nu}(gk_1)^\mu
(gk_2)^\nu}\,a^\dagger_{gk_2}a^\dagger_{gk_1}|0\rangle \ .
\end{equation}

This can be achieved if $U(g)$ and $a^\dagger_k$ satisfy
\begin{equation}
U(g)a_k^\dagger=a_{gk}^\dagger
e^{-\frac{i}{2}\theta_{\mu\nu}(gk)^\mu P^\nu}
e^{\frac{i}{2}\theta_{\mu\nu}(k)^\mu (g^{-1})^\nu_\rho P^\rho}U(g)\
,\label{adtrans}
\end{equation}
where $g$ on the right hand side stands for the matrix of the
Lorentz transformation and
\begin{equation}
P^\mu=\int \frac{d^3k}{2k^0} k^\mu a_k^\dagger a_k \ .
\end{equation}
Notice that just like the usual momentum operator, $P^\mu$ obeys
\begin{align}
[P^\mu , a^\dagger_k]&=k^\mu a^\dagger_k \ ,\\
[P^\mu, P^\nu] &= 0\ .
\end{align}
Hence we have
\begin{align}
U(g)a^\dagger_{k_2}&a^\dagger_{k_1}|0\rangle \nonumber\\
&=a_{gk_2}^\dagger e^{-\frac{i}{2}\theta_{\mu\nu}(gk_2)^\mu P^\nu}
e^{\frac{i}{2}\theta_{\mu\nu}k_2^\mu ((g^{-1})^\nu_\rho P^\rho)}
a_{gk_1}^\dagger e^{-\frac{i}{2}\theta_{\mu\nu}(gk_1)^\mu P^\nu}
e^{\frac{i}{2}\theta_{\mu\nu}k_1^\mu ((g^{-1})^\nu_\rho
P^\rho)}U(g)|0\rangle \nonumber\\
&=a_{gk_2}^\dagger a_{gk_1}^\dagger
e^{-\frac{i}{2}\theta_{\mu\nu}(gk_2)^\mu (gk_1)^\nu}
e^{\frac{i}{2}\theta_{\mu\nu}k_2^\mu ((g^{-1})^\nu_\rho
g^\rho_\sigma
k_1^\sigma)}|0\rangle \nonumber\\
&=e^{\frac{i}{2}\theta_{\mu\nu}(gk_1)^\mu (gk_2)^\nu}
e^{-\frac{i}{2}\theta_{\mu\nu}k_1^\mu k_2^\nu}a_{gk_2}^\dagger
a_{gk_1}^\dagger|0\rangle
\end{align}
as required.

This works for any number of particles. We can write
(\ref{adtrans}) as
\begin{equation}
U(g)a^\dagger_k=\rho_{k'k}(g_\alpha^{(2)})a_{k'}^\dagger
U(g_\alpha^{(1)})
\end{equation}
where $g^{(i)}_\alpha$ defined by coproduct (cf.
Eq.(\ref{sweedler})).
 Note the ordering of $g^{(2)}$ and $g^{(1)}$. Now we have
\begin{align}
U(g)a^\dagger_{k_n}&a^\dagger_{k_{n-1}}\cdots a^\dagger_{k_1}|0\rangle  \nonumber\\
=&\rho_{k_n'k_n}(g^{(2)})\rho_{k_{n-1}'k_{n-1}}(g^{(1)(2)})
\cdots\rho_{k_1'k_1}(g^{(1)(1)\cdots(2)})\nonumber\\
&a^\dagger_{k'_n}a^\dagger_{k'{n_1}}\cdots a^\dagger_{k_1}U(g^{(1)(1)\cdots(1)})|0\rangle \nonumber\\
=&\rho_{k_n'k_n}(g^{(2)})\rho_{k_{n-1}'k_{n-1}}(g^{(1)(2)})\cdots
\rho_{k_1'k_1}(g^{(1)(1)\cdots(2)})a^\dagger_{k'_n}a^\dagger_{k'_{n-1}}\cdots a^\dagger_{k'_1}\epsilon(g^{(1)(1)\cdots(1)})|0\rangle \nonumber\\
=&a^\dagger_{k'_n}a^\dagger_{k'_{n-1}}\cdots
a^\dagger_{k_1}(id\otimes\rho_{k_1'k_1}\otimes\rho_{k_2'k_2}\otimes\cdots\otimes\rho_{k_n'k_n})\nonumber\\
&(\epsilon\otimes id\otimes id\otimes\cdots\otimes
id)(g^{(1)(1)\cdots(1)}\otimes g^{(1)(1)\cdots(2)}\otimes
g^{(2)})|0\rangle \nonumber\\
=&a^\dagger_{k'_n}a^\dagger_{k'_{n-1}}\cdots a^\dagger_{k'_1}
(\rho_{k_1'k_1}\otimes\rho_{k_2'k_2}\otimes\cdots\otimes\rho_{k_n'k_n})\nonumber\\
&(g^{(1)(1)\cdots(2)}\otimes g^{(1)(1)\cdots(2)}\cdots\otimes g^{(2)}|0\rangle \nonumber\\
=&\rho_{k_1'k_1}\,\otimes\,\rho_{k_2'k_2}\,\otimes\,\cdots\,\otimes\,\rho_{k_n'k_n}\,\nonumber\\
&\{(\mathds{1}\,\otimes\,\cdots\,\otimes\,\mathds{1}\,\otimes\,\Delta_\theta)\,\cdots\,
(\mathds{1}\,\otimes\,\Delta_\theta)\,\Delta_\theta(g)\}\,|k_1'\,,\,k_2'\,,\,\cdots\,k_n'\rangle_{s_\theta}
\end{align}
which is as desired. Here we have used
\begin{equation}
(\epsilon\otimes id)\Delta=id\label{eq}
\end{equation}
where $\epsilon$ is the counit.

We can write eq (\ref{adtrans}) in a covariant way as
(cf.(\ref{sweedler}))
\begin{equation}
U(g)a^\dagger_k=a^\dagger_{k'}(U\otimes \rho_{k'k})\Delta(g)\ \
\end{equation}
or
\begin{equation}
U(g)a^\dagger_k=a^\dagger_{k'}(\rho_{k'k}\otimes U)\Delta'(g)\ .
\end{equation}

 \subsection{The Quantum Operators for the Poincar\'e group}
 Now we discuss the representation, $U(g)$, of the generators of the twisted
 Poincar\'e group on the Hilbert space of the theory. These generators must transform the
 multiparticle states according to the twisted coproduct and hence
 fulfil the eq.(\ref{adtrans}).

As discussed in \cite{bal}  we have a map, ``dressing
transformation''\cite{grosse,Zamolodchikov:1978xm,Faddeev:1980zy},
between usual creation/annihilation operators
 and the twisted ones given by
 \begin{align}
 a^\dagger_k&=c^\dagger_k e^{\frac{i}{2}\theta_{\mu\nu}k^\mu
 P^\nu}\ ,\nonumber\\
 a_k&=c_k e^{-\frac{i}{2}\theta_{\mu\nu}k^\mu P^\nu}\ ,\label{map}
 \end{align}
 where $c_k$ and $c^\dagger_k$ have the usual bosonic commutation
 relations. Note that the momentum operator has the same expression
 in terms of $a_k$'s and $c_k$'s :
 \begin{equation}
 P^\mu=\int \frac{d^3k}{2k^0} k^\mu a_k^\dagger a_k
 =\int \frac{d^3k}{2k^0} k^\mu c_k^\dagger c_k \ .\label{mom}
 \end{equation}
 This is as expected since the twist does not change the coproduct of the
 momentum.

 With the above expressions, $a^\dagger_k$ and $a_k$ satisfy the
 correct commutation relations (\ref{adcom}),(\ref{com}),(\ref{adacom}), and the states created by
 $a^\dagger_k$ have the correct scalar product.
 Hence, for any operator, we can replace $a_k$'s and $a^\dagger_k$'s by
 the expressions in (\ref{map}), to express it in terms of regular
 creation and annihilation operators $c_k$ and $c^\dagger_k$'s.

To find the representation of the twisted Poincar\'e group, we
will use as a guiding principle the result of
\cite{Balachandran:2006ib,Balachandran:2007kv}, that one can use a
representation of the commutative algebra on the noncommutative
one to construct a representation of the twisted Poincar\'e group.
The correct representation is just the usual one written in terms
of the commutative algebra.\footnote{Note that for the generator
of the momentum, this is trivially true, cf. eq.(\ref{mom})} We
will prove that the same is true in quantum case too. Namely, we
will take as the representation the usual expressions for the {\it
untwisted} Poincar\'{e} generators written in terms of $c_k$ and
$c^\dagger_k$'s. It is a trivial observation that they will
satisfy the standard commutation relations as required. So the
group structure  is correct. Now we show that the same is true for
the coproduct.

As  was shown in the previous section, all we need to show is that
$U(g)$'s acting on $a_k^\dagger$ satisfy (\ref{adtrans}). Since
$U(g)$ is an untwisted generator, $c^\dagger_k$ transforms under
its action in the usual way (cf. (\ref{adtrans}) for
$\theta_{\mu\nu}=0$):
 \begin{equation}
 U(g)c^\dagger_k=c^\dagger_{gk}U(g)\ .
 \end{equation}
$P^\mu$, as all generators, has the usual expression in terms of
$c^\dagger_k,c_k$'s. Hence we have
 \begin{equation}
 U(g)P^\mu=(g^{-1})^\mu_\nu P^\nu U(g) \ .
 \end{equation}
 Using this, we have
 \begin{align}
 U(g)a_k^\dagger=& U(g)c^\dagger_k e^{\frac{i}{2}\theta_{\mu\nu}k^\mu
 P^\nu}\nonumber\\
 =& c^\dagger_{gk}e^{\frac{i}{2}\theta_{\mu\nu}k^\mu
 ((g^{-1})^\nu_\rho P^\rho)}U(g)\nonumber \\
 =& a^\dagger_{gk} e^{-\frac{i}{2}\theta_{\mu\nu}(gk)^\mu
 P^\nu} e^{\frac{i}{2}\theta_{\mu\nu}k^\mu
 ((g^{-1})^\nu_\rho P^\rho)}U(g)\ ,
 \end{align}
 which fulfills Eq.(\ref{adtrans}).

This completes the proof that the usual representation of the
Poincar\'{e} generators, constructed out of untwisted
creation/annihilation operators, $c_k$ and $c^\dagger_k$, acting
on the twisted Hilbert space realizes the representation of the
{\it twisted} Poincar\'{e} group. This allows the discussion of
twisted symmetries in a very simple and general way. Using this
technique one can easily infer the transformation properties under
C, P and T etc. Perhaps the full diffeomorphism symmetry can also
be addressed in this manner.

 As an illustration, consider the two particle state
 \begin{equation}
 |k_1,k_2\rangle_{s_\theta} =a_{k_2}^\dagger a_{k_1}^\dagger|0\rangle =e^{\frac{i}{2}\theta_{\mu\nu}k_2^\mu k_1^\nu} c_{k_2}^\dagger
c_{k_1}^\dagger|0\rangle\ .
\end{equation}
Now acting by $U(g)$, we just transform the $c^\dagger$'s to get,
\begin{align}
U(g)|k_1,k_2\rangle_{s_\theta}  &=
e^{-\frac{i}{2}\theta_{\mu\nu}k_1^\mu k_2^\nu}
c_{gk_2}^\dagger c_{gk_1}^\dagger|0\rangle \\
&=e^{-\frac{i}{2}\theta_{\mu\nu}k_1^\mu
k_2^\nu}e^{+\frac{i}{2}\theta_{\mu\nu}gk_1^\mu gk_2^\nu}
|gk_1,gk_2\rangle_{s_\theta} \\
 &=\rho_{k_1'k_1}\otimes \rho_{k_2'k_2}
(\Delta_\theta(g))|k_1',k_2'\rangle_{s_\theta}\ .
\end{align}
as required.

\section{On Invariant Interactions}\label{ih}

The complete treatment of the construction of the invariant
$S$-matrix is given in
\cite{Balachandran:2006ib,Balachandran:2007yf}. There the precise
conditions on the Hamiltonian are derived. Here we demonstrate how
elementary is the proof of the uniqueness of the form of the
Hamiltonian (in the sense that all fields should be multiplied
with star-products) if one uses our result from the previous
section on the representation of the twisted Poincar\'{e} group.

\subsection{Fields}
Let us begin with the introduction of quantum fields. This is done
in complete analogy with the usual case. We define creation and
annihilation fields from the creation and annihilation operators,
\begin{align}
\Phi^+(x)&=\int d^3k e^{ik\cdot x}a_k^\dagger\ , \nonumber\\
\Phi^-(x)&=\int d^3k e^{-ik\cdot x}a_k \label{fields}
\end{align}
and the real Hermitian field
\begin{equation}
\Phi(x)=\Phi^+(x)+\Phi^-(x)\ .\label{field}
\end{equation}
Using eq  (\ref{adtrans}), we find that the field $\Phi(x)$ obeys
the following transformation law
\begin{equation}
U(g)\Phi(x)=\Phi(gx) e^{-\frac{1}{2}\theta_{\mu\nu}(g^\mu_\sigma
{\overleftarrow \partial}^\sigma) P^\nu}
e^{\frac{1}{2}\theta_{\mu\nu}{\overleftarrow \partial}^\mu
(g^{-1})^\nu_\rho P^\rho}U(g)\ ,\label{fldtrans}
\end{equation}
where ${\overleftarrow \partial}^\sigma$ acts to the left, only on
the field argument, while $P^\mu$ is the total momentum operator
and acts on everything to the right. But what is more important,
the twisted field $\Phi(x)$, (\ref{field}), can be written in
terms of untwisted field $\Phi_0(x)$, constructed as in
(\ref{fields}) but with $c_k^\dagger$ and $c_k$. Using
(\ref{map}), we easily find
\begin{equation}
\Phi(x) = \Phi_0(x) e^{\frac{1}{2}
{\overleftarrow\partial}^\mu\theta_{\mu\nu} P^\nu}\ .
\end{equation}
Using this result, we see that the following is true
\begin{equation}
\Phi(x)*\Phi(x) = \left(\Phi_0(x)\Phi_0(x)\right)e^{\frac{1}{2}
{\overleftarrow\partial}^\mu\theta_{\mu\nu} P^\nu}\ \label{trivial}
\end{equation}
where $*$ was defined in (\ref{star}). This is a very important
result, which makes the proof of the above statement about the
interaction Hamiltonian almost trivial as we now demonstrate.

\subsection{Interaction Hamiltonian}

The first non-trivial term in the expansion of the $S$-matrix is
\begin{equation}
S^{(1)}=\int d^4x\,\mathcal{H_I}(x)\ .
\end{equation}
As we want to have a twisted invariant $S$-matrix, this term should
be invariant separately. We claim that if $\mathcal{H_I}(x)$ is a
star-polynomial in $\Phi(x)$ with a typical term being
\begin{equation}
\mathcal{H}_I(x)=\Phi(x)*\Phi(x)*\cdots *\Phi(x)
\end{equation}
then $S^{(1)}$ is twisted invariant. The proof is the combination of
(\ref{trivial}) and our construction of the generators of the
twisted symmetry.

From (\ref{trivial}) we have
\begin{equation}
\Phi(x)^n_* = \left(\Phi_0(x)^n\right)e^{\frac{1}{2}
{\overleftarrow\partial}^\mu\theta_{\mu\nu} P^\nu}\
,\label{trivial1}
\end{equation}
so the integration over 4-d (assuming that fields behave "nicely"
at infinity) gives
\begin{equation}
\int d^4x\,\Phi(x)^n_* = \int
d^4x\,\left(\Phi_0(x)^n\right)e^{\frac{1}{2}
{\overleftarrow\partial}^\mu\theta_{\mu\nu} P^\nu} = \int
d^4x\,\Phi_0(x)^n\ ,
\end{equation}
i.e., just a commutative result. Now we use the fact that the
generators of the twisted symmetry are represented by the usual
commutative generators constructed out of $c_k^\dagger$ and $c_k$
(or equivalently, out of $\Phi_0(x)$). Then the invariance follows
immediately.

By similar calculations, it is elsewhere shown to all orders that
the $S$-operator is independent of $\theta_{\mu\nu}$
\cite{stat,oeckl,Balachandran:2006ib,Balachandran:2007kv,Balachandran:2007yf,Balachandran:2005pn}.

As the $S$-operator is the same as the $S$-operator of the
commutative theory (though the $S$-matrix is not), all trace of
$\theta_{\mu\nu}$ is only in the statistics of in and out states.

It is worth noting that the whole structure developed so far is very
rigid, in particular the $S$-operator is the only one which commutes
with the symmetry operators giving us a twisted Poincar\'e invariant
$S$-matrix.

All these considerations are valid only in the absence of gauge
fields. Gauge field theories in our approach are discussed in
\cite{Balachandran:2006ib,Balachandran:2007kv,Balachandran:2007yf,Akofor:2007hk}.

\section{On the Definition of Creation Operators}\label{sdef}
Here we comment on the ambiguity in the definition of
creation/annihilation operators. As we saw in the section
\ref{qft}, we have to make a particular choice in the definition
of  creation operators, as to whether they add a particle to the
left or to the right of the list of particles already present in
the state. By now it should be clear  what would happen if we had
defined creation operators as adding the particle to the left of
the list. The transformation properties and the statistics of the
states of course do not depend on it as it was already fixed by
the structure of the group. But the transformation of  creation
operators themselves and their commutation relations will be
changed . Now they look like
\begin{align}
U(g)a^\dagger_k&=a^\dagger_{k'}(\rho_{k'k}\otimes U )\Delta(g),\\
a_{k_2}^\dagger a_{k_1}^\dagger\ &=\ e^{-i\theta_{\mu\nu}k_2^\mu
k_1^\nu}\,a_{k_1}^\dagger a_{k_2}^\dagger\ .
\end{align}
As a consequence, the invariant Hamiltonian now must be defined with
a $\ast^{-1}$-product i.e., with a product defined by
\begin{equation}
\Phi(x)\ast^{-1}\Phi(x)=m_0
\mathcal{F}_{\theta}^{-1}\Phi(x)\otimes\Phi(y)\ .
\end{equation}
That is why we adopted the other choice, as it gives the familiar
$\ast$-product in the Hamiltonian. But the result that the
$S$-operator is independent of $\theta_{\mu\nu}$ still holds. Hence
this choice does not make any physical difference as the $S$-matrix
of the theory is the same in both cases. In both cases , finally we
have a theory with the same statistics for asymptotic states and the
usual $S$-operator.

\section{Conclusions}\label{scon}

We have studied the general construction of field theories with
twisted Poincar\'e invariance. A necessary consequence of such a
symmetry is the twisting of statistics. The requirement of
invariance of the $S$-matrix (and a convention regarding the
definition of twisted creation and annihilation operators) forces
one to choose the $\ast$-product between the fields in the
interaction Hamiltonian. The final theory, in the most general
case, consists of a theory with twisted statistics of the
asymptotic states and an $S$-operator which is completely
independent of the noncommutativity parameter $\theta_{\mu\nu}$.

 \acknowledgments

The work of APB and BAQ is supported by the D.O.E. grant number
DE-FG02-85ER40231. The work of AP is supported by FAPESP grant
number 06/56056-0.

\end{document}